\documentclass[12pt,a4paper]{article}
\usepackage[paperwidth=27cm, paperheight=33cm]{geometry}
\usepackage[T1]{fontenc}
\usepackage[latin1]{inputenc}
\usepackage{verbatim,epstopdf,epsfig,graphicx,amsthm,amsmath,amssymb,amsthm,amsfonts,titling,url,array,subfig}
\usepackage[nottoc]{tocbibind}
\usepackage[toc,page]{appendix}
\usepackage{slashed}

\title{\textbf{Integrability of the Dirac Equation in the Presence of Fluxes on Product Manifolds}}
\author{Jo\'{a}s Ven\^{a}ncio and Azadeh Mohammadi }
\date{}

\begin{document}
\maketitle
\vspace{-2.4cm}
\begin{flushleft}
\textit{Departamento de F\'{\i}sica, Universidade Federal de Pernambuco,
Recife, Pernambuco  50740-560, Brazil}
\end{flushleft}

\vspace{.5cm}

This paper aims to show that the Dirac equation coupled to an arbitrary inhomogeneous flux field admits separation in manifolds formed from the direct product of bidimensional spaces. As a direct application of these results, we study a spin-$1/2$ charged particle propagating in a background conformally related to a novel and complex string-inspired model in $D = 10$ spacetime dimension whose base manifold is a product of four two-dimensional unit spheres, $S^{2}$, in the presence of $1$- and $3$-form fluxes.

\vspace{.3cm}

\section{Introduction}

In any theory of gravity, an efficient and accurate way to probe the gravitational field permeating a given space ($M, \boldsymbol{g}$) endowed with a metric $\boldsymbol{g}$ is by letting other fields interact with it. This makes studies of the behavior of a matter field (scalar, spinorial, or abelian and non-abelian gauge fields) on ($M, \boldsymbol{g}$) a central piece of investigation in general relativity. In order to solve the field equation for a given matter field propagating in a given space, the first step is to separate the degrees of freedom of the field by carefully choosing a suitable basis that allows us to decouple the variables in the relevant field equation. However, the problem of integrating and separating the matter field equation on ($M, \boldsymbol{g}$) is highly non-trivial. In general, the variables in matter field equations cannot be decoupled for fields propagating on an arbitrary space once an arbitrary space possesses no symmetry. Indeed, the choice of a suitable basis depends directly on the symmetries of the space. So, for the integration process to be possible, the space must possess sufficient symmetry. It is possible to reconcile the separability properties of these matter field equations with the existence of certain symmetry tensors, such as Killing tensors, Killing-Yano tensors, and their conformal versions \cite{Carter1968,Hughston1973,Frolov2008,Carlos2014}. Besides that, when one writes a matter field equation on a given spacetime, the parameters that describe it uniquely, such as mass, electric and magnetic (if such exists) charges, and angular momentum, become parameters of the field equation. This paper concentrates on the spinor field.

Spinors have been around for a long time. They were first discovered a century ago by {\'E}lie Cartan \cite{Cartanbook66} and play a central role in the theory of particles and fields, which are represented by spinor fields or by their tensor products. Spinors are objects that provide the least-dimensional faithful representations for the group $Spin(D)$, the universal covering of the group $SO(D)$, the rotation group in $D$ dimensions \cite{Cartanbook66,Cartan1913}. Moreover, the Lorentz group can be seen as the rotation group in a space with a  Lorentzian signature metric so that spinors can be the building block of all other representations of this group. In this sense, spinors are the most fundamental objects of a space endowed with a metric.  
Besides that, a specific class of spinor fields plays an important role when dealing with supersymmetric theories, the so-called Killing spinor fields. Through the bilinears constructed from a single Killing spinor field, generating a whole tower of Killing-Yano tensors is possible. This geometric structure can generate extra supersymmetries in the phase space, as illustrated by the theory of the supersymmetric spinning particle \cite{Gibbons1993,Cariglia2004}. Killng spinors can also be used to investigate whether a given solution can be supersymmetric as well as determine the number of supersymmetries \cite{Nieuwenhuizen1984}.
In order to gain a better understanding of spinor techniques, it is interesting to investigate spinor fields propagating in more general curved spacetimes in dimensions greater than four. Among the vast literature concerning the propagation of matter fields in curved spacetimes, there are only a few works concerning the spinor fields. The reason for this is not conceptual but the more significant technical complexity involved in the description of spinor fields, since the spinor formalism takes fully into account the dimensional specificities, in general \cite{Benn_Book1987,Joas_Book2019,Joas_Formalism,Carlos_Classification}. This paper is also an effort to fill a part of this shortfall.

Spinor fields satisfying the standard Dirac equation represent spin-1/2 particles, such as protons, electrons, and neutrons. Since its derivation in 1928, stimulated by the progress in general relativity and quantum field theory, the Dirac equation and its separability properties have been investigated in increasingly complicated curved spacetimes. The separability of the massless Dirac equation on Schwarzschild spacetime was discussed first by Brill and Wheeler \cite{Brill1957}, and by Unruh on Kerr spacetime \cite{Unruh:1973bda}, and then extended for the massive case by Chandrasekhar in a remarkable work \cite{Chandrasekhar1979} using the Newman-Penrose formalism \cite{Newman1962}. The Chandrasekhar result was generalized to a Kerr-Newman black hole case by Page \cite{Page:1976jj}. Many of these results were summarized by Teukolsky into a master equation, which furnishes the separability for equations of motion for fields with an arbitrary spin on Kerr spacetime \cite{Carter1968,Walk-Pen1970}. Wu \cite{Wu2009} studied the separability of the Dirac equation on a generalization of the famous four-dimensional Kerr-Newman solution,  a five-dimensional rotating, charged black hole constructed by Cveti$\check{\text{c}}$ and Youm in string theory. Some classes of black holes can also have magnetic charges, called the dyonic case. Semiz \cite{Semiz1992}, for instance, carried out the separability of the Dirac equation on a dyonic Kerr-Newman black hole, while Dudley and Finley \cite{Dudley1979} investigated the separability of massless field equation for any spin in a class of dyonic Petrov type-D solutions. 
We have given some contributions in this respect in \cite{Joas2017, Joas2018}, in which we work out the separation of the massive Dirac equation minimally coupled to the background gauge field on a higher-dimensional generalization of the charged Schwarzschild and Nariai spacetimes, both having magnetic charges.

The separation of variables in the Dirac equation, in general, is a challenging task. The main reason is that while for the geodesic Hamilton-Jacobi and the Klein-Gordon scalar field equations, there exists a well-understood separability theory \cite{Carter1968,Benenti1997,Demianski1980,Fatibene2002}, for the Dirac equation such a complete separability theory is still lacking. Since there are no general methods to attain Dirac's equation separability, this mathematical problem must be carried out separately in each case. However, the Dirac equation once seperated in variables and solved, one can study many interesting phenomena, including fermionic vacuum polarization, Casimir effect, induced currents in nontrivial background spacetimes, topological defects for instance \cite{Mohammadi2013,Mohammadi2015,Mohammadi2020}.
The separability properties of the Dirac equation for some cases are shown to be closely connected with the existence of certain symmetry operators, generated by Killing-Yano tensors, that commute with the Dirac operator and characterize the separation constants involved in the problems \cite{Charlton1997, Carter1979}. In the Kerr geometry, Carter and McLenaghan \cite{Carter1968,Carter1979} found that the separability of the Dirac equation obtained by Chandrasekhar \cite{Chandrasekhar1979} is related to the fact that the Kerr geometry admits a geometrical structure called Killing-Yano tensor that allows the integration of this field equation. Cariglia \cite{Cariglia2011} characterized the separability of the Dirac equation in the background of Kerr-NUT-(A)dS spacetimes in arbitrary dimensions and made the connection with the general solution of the Dirac equation found by Oota and Yasui \cite{Oota2008}. However, in higher dimensions, there is not even a general understanding of how the existence of these symmetry operators relates to the geometrical structures in the space on which the Dirac equation is defined, that is, the necessary and sufficient conditions for separability.

Several theories attempt to extend our understanding of the Universe through higher dimensional theories, especially after the observational discovery of the expansion of the Universe. The most popular one is the (super)string theory, an elegant type of quantum field theory based on the implementation of supersymmetry as gauge symmetry \cite{Freedman2012}. In our previous work \cite{Joas2017}, we investigated the separability of a Dirac equation coupled to a 1-form field (gauge field) on a higher dimensional class that is the direct product of bidimensional spaces. Particularly, spaces that are the direct product can also be relevant to model base spaces in string theory compactifications \cite{Brown2014B,Brown2014A}.
In these higher dimensional theories, besides higher-order terms in curvature, the metric is often supplemented by other fields that couple to the spinor fields and modify the standard Dirac equation. These other fields are usually called fluxes \cite{Martucci2005,Tripathy2005,Grana2002}. It is therefore,  quite natural to investigate the separability of the Dirac equation for the case when the spaces admitting these fluxes \cite{Cariglia2012,Kubiznak2011}.

Motivated by ideas arising in some models inspired by string theory \cite{Giribet2018}, this paper will specifically focus on the separability of the Dirac equation in the presence of arbitrary fluxes in the same class of spaces studied in previous works \cite{Joas2017,Joas2018}, comprised of the direct product of two-dimensional spaces. As a direct application of these results, we shall consider the Dirac equation for a massive charged spin-$1/2$ particle propagating on a background conformally related to complex string-inspired models in $D = 10$ spacetime dimension, a black hole solution charged under both the $1$- and $3$-form fields. The base manifold is a direct product of two-dimensional unit spheres, described in Ref \cite{Giribet2018}. A curious point about this black hole class is that when a quadratic-order term in curvature and a fourth-order term in $4$-form flux is included in the action, it is possible to write a dyonic black hole solution in a closed form in the general relativity limit. This provides a tractable model with rich physics. Besides that, this black hole supports magnetic charges besides the electric charge, which leads to richer physics. 


The organization of this paper is outlined as follows. Section \ref{Product_Manifolds} sets the necessary tools to be used throughout the paper in order to attain the separability of the Dirac equation with fluxes on the background of spaces constituting the direct product of two-dimensional spaces. In Sec. \ref{Separability_3D}, we focus on the separability of the Dirac equation in this background for the case where the fluxes consist of a $p$-form where $p \leq 3$. Then, in Sec. \ref{Separability_GeneralCase}, we generalize the previous results for an arbitrary inhomogeneous form flux. Section \ref{String-Inspired} uses the results obtained in the previous sections to attain the separation of the Dirac equation on a background conformally related to a dyonic black hole in $D=10$ spacetime dimensions. Finally, Sec. \ref{Conclusion} ends with a brief summary of this paper and comment on future applications of this research.

\section{Product Manifolds and Dirac Equation with Fluxes}\label{Product_Manifolds}

The present section aims to set up the scenario to show that the Dirac equation with flux is separable in spaces that are the direct product of bidimensional spaces. This class of spaces are manifolds $(M, \hat{\boldsymbol{g}\,})$ which can be equipped with a metric tensor $\hat{\boldsymbol{g}\,}$ whose components with respect to an orthonormal basis $\{\hat{\boldsymbol{X}\,}_\alpha\}$ for $TM$ are given by
\begin{eqnarray}\label{IPVF}
\hat{\boldsymbol{g}\,}(\hat{\boldsymbol{X}\,}_{\alpha}, \hat{\boldsymbol{X}\,}_{\beta}) \,=\, \delta_{\alpha\beta} \quad \leftrightarrow  \quad \left\{\begin{matrix}
\hat{\boldsymbol{g}\,}(\hat{\boldsymbol{X}\,}_{a}, \hat{\boldsymbol{X}\,}_{b}) &=& \delta_{ab},\\ 
\hat{\boldsymbol{g}\,}(\hat{\boldsymbol{X}\,}_{a}, \hat{\boldsymbol{X}\,}_{\widetilde{b}}) &=& 0,\\ 
\hat{\boldsymbol{g}\,}(\hat{\boldsymbol{X}\,}_{\widetilde{a}}, \hat{\boldsymbol{X}\,}_{\widetilde{b}}) &=& \delta_{\widetilde{a}\widetilde{b}},
\end{matrix}\right.
\end{eqnarray}
where the indices $a$ and $b$ label the first $n$ vector fields of $\{\hat{\boldsymbol{X}\,}_\alpha\}$, while $\widetilde{a}$ and $\widetilde{b}$ label the remaining $n$ vector fields of $\{\hat{\boldsymbol{X}\,}_\alpha\}$. The manifold $(M, \hat{\boldsymbol{g}\,})$ is a direct product of $n$ bidimensional spaces which can be covered by coordinates
$\{x^{1} , y^{1} , x^{2} , y^{2} , \ldots , x^{n}, y^{n}\}$ such that the line element is written as
\begin{eqnarray}\label{2-dimens.spaces}
d\hat{s}^{2}\,=\,\sum_{a=1}^{n}\,d\hat{s}_{a}^{2}\,=\,\sum_{a=1}^{n}\,(\hat{\boldsymbol{e}}^{a}\,\hat{\boldsymbol{e}}^{a}\,+\,\hat{\boldsymbol{e}}^{\widetilde{a}}\,\hat{\boldsymbol{e}}^{\widetilde{a}}) \,,
\end{eqnarray}
where $d\hat{s}_{a}^{2}$ are two-dimensional line elements depending just on the two coordinates $\{x^{a}, y^{a}\}$, corresponding to their bidimensional spaces, and the $1$-forms $\{\hat{\boldsymbol{e}}^{\alpha}\}$ provide a basis for $T^{*}M$, the dual of the tangent space $TM$.   
Derivatives of the vector fields determine the spin connection according to the following relation
\begin{equation}\label{covD}
\hat{\nabla}_{\alpha} \hat{\boldsymbol{X}\,}_{\beta} \,=\, \hat{\omega}_{\alpha\beta}^{\phantom{\alpha\beta}\gamma}\,\hat{\boldsymbol{X}\,}_{\gamma} \,.
\end{equation}
Since the metric $\hat{\boldsymbol{g}\,}$ is a covariant constant tensor, it follows that the coefficients of the spin connection with all lower indices $\hat{\omega}_{\alpha\beta\gamma} = \hat{\omega}_{\alpha\beta}^{\phantom{\alpha\beta}\varepsilon}\,\delta_{\varepsilon\gamma}$ are antisymmetric in their two last indices, $\hat{\omega}_{\alpha\beta\gamma} = - \hat{\omega}_{\alpha\gamma\beta}$. However, since the indices of the spin connection are lowered and raised with $\delta_{\alpha\beta}$ and  $\delta^{\alpha\beta}$, respectively, the frame indices can be raised and lowered unpunished. In particular, with respect to the orthonormal basis of vectors \eqref{IPVF}, all the nonvanishing components of the spin connection are the following
\begin{eqnarray}\label{spin-connec}
\hat{\omega}_{aa\widetilde{a}} \,=\, -\,\hat{\omega}_{a\widetilde{a}a}, \qquad 
\hat{\omega}_{\,\widetilde{a}a\widetilde{a}} \,=\, -\,\hat{\omega}_{\,\widetilde{a}\widetilde{a}a}.
\end{eqnarray}

The vector fields $\hat{\boldsymbol{X}\,}_{\alpha}$ can be represented by Dirac matrices $\gamma_{\alpha}$ which, in $2n$ dimensions, represent faithfully the Clifford algebra by $2^{n} \times 2^{n}$ matrices obeying the following anticommutation relation
\begin{equation}\label{Clifford-Algebra}
   \gamma_{\alpha} \gamma_{\beta} + \gamma_{\beta}\gamma_{\alpha} =2 \,  \hat{\boldsymbol{g}\,}(\hat{\boldsymbol{X}\,}_\alpha,\hat{\boldsymbol{X}\,}_\beta)\,\mathbb{I}_n\,,
\end{equation}
with $\mathbb{I}_n$ standing for the $2^{n}\times 2^{n}$ identity matrix.

In order to accomplish the separability of the Dirac equation, let us introduce a convenient representation for the Dirac matrices. In terms of the Pauli matrices 
\begin{equation}
\sigma_1 = \left[
                  \begin{array}{cc}
                    0 & 1 \\
                    1 & 0 \\
                  \end{array}
                \right] \, , \;\;
                \sigma_2 = \left[
                  \begin{array}{cc}
                    0 & -i \\
                    i & 0 \\
                  \end{array}
                \right]  \, , \;\;
                \sigma_3 = \left[
                  \begin{array}{cc}
                    1 & 0 \\
                    0 & -1 \\
                  \end{array}
                \right]\,,
\end{equation}
such a representation can be expressed in $2n$ dimensions as
\begin{align}
  \gamma_{a} & = \underbrace{\sigma_3\otimes \cdots \otimes  \sigma_3}_{(a-1)\;\textrm{times}}  \otimes  \,\sigma_1  \otimes
  \underbrace{\mathbb{I} \otimes \cdots \otimes  \mathbb{I}}_{(n-a)\;\textrm{times}} \,,\nonumber \\
  \quad \label{DiracMatrices}\\
  \gamma_{\widetilde{a}} & = \underbrace{\sigma_3\otimes \cdots \otimes  \sigma_3}_{(a-1)\;\textrm{times}}  \otimes  \,\sigma_2  \otimes
  \underbrace{\mathbb{I} \otimes \cdots \otimes  \mathbb{I}}_{(n-a)\;\textrm{times}} \,,\nonumber
\end{align}
where the index $a$ ranges from $1$ to $n$ and $\mathbb{I}$ is the $2\times 2$ identity matrix. Indeed, we can easily check that the Clifford algebra given in Eq. \eqref{Clifford-Algebra} is properly satisfied by the above matrices. In this case, the spinor space $SM$ can be thought of as being the space of column vectors with $n$ entries on which these matrices act, and its elements are called spinor fields, denoted by $\hat{\boldsymbol{\Psi}}$.

A basis for the spinor space $SM$ can be spanned by direct products of the following spinors in two dimensions
\begin{equation}\label{SpinBasis}
  \boldsymbol{\xi}^+ = \left[
              \begin{array}{c}
                1 \\
                0 \\
              \end{array}
            \right] \quad \textrm{and} \quad
             \boldsymbol{\xi}^- = \left[
              \begin{array}{c}
                0 \\
                1 \\
              \end{array}
            \right].
\end{equation}
Under the action of the Pauli matrices, the above spinors satisfy concisely the relations
\begin{equation}\label{PauliAction}
  \sigma_1\boldsymbol{\xi}^{s_{a}} =  \boldsymbol{\xi}^{-{s_{a}}}, \qquad  \sigma_2\boldsymbol{\xi}^{s_{a}} = i\,s_{a}\,\boldsymbol{\xi}^{-s_a} , \qquad  \sigma_3\boldsymbol{\xi}^{s_{a}} = s_{a}\,\boldsymbol{\xi}^{s_{a}} \quad (\text{no sum over $a$}),
\end{equation}
where the spinor index $s_a$ can take the values ``+1''
and ``-1''. In terms of these column vectors, any spinor field in $2n$ dimensions can be written as
\begin{equation}\label{SpinBasis2}
\hat{\boldsymbol{\Psi}} \,=\,\sum_{\{s\}}\hat{\Psi}^{s_{1}s_{2}\dots s_{n}}\boldsymbol{\xi}^{s_{1}}\otimes \boldsymbol{\xi}^{s_{2}} \otimes \dots \otimes \boldsymbol{\xi}^{s_{n}}.
\end{equation}
Indeed, since each of the indices $s_{a}$ can take only two values, it follows the sum over $\{s\} \equiv \{s_1, s_2,\ldots, s_n \}$ comprises $2^n$ terms, which is precisely the dimension of the spinor space in $2n$ dimensions. 

By construction, the action of $\gamma_{\alpha}$ on $SM$ yields elements on $SM$. Indeed, using the equations (\ref{DiracMatrices}), (\ref{PauliAction}) and (\ref{SpinBasis2}), we eventually arrive at the following identity
\begin{align}\label{GammaaPsi}
 \gamma_a \hat{\boldsymbol{\Psi}} & = \sum_{\{s\}}\left(\prod_{b=1}^{a-1}s_b\right) \hat{\Psi}^{s_1s_2\cdots s_{a-1} s_a s_{a+1}\cdots s_n} \;
\boldsymbol{\xi}^{s_1}\otimes \boldsymbol{\xi}^{s_2}\otimes \cdots \otimes \boldsymbol{\xi}^{s_{a-1}}\otimes  \boldsymbol{\xi}^{-s_{a}} \otimes \boldsymbol{\xi}^{s_{a+1}} \otimes  \cdots \otimes  \boldsymbol{\xi}^{s_n} \nonumber\\
& =  \sum_{\{s\}}\left(\prod_{b=1}^{a-1}s_b\right) \hat{\Psi}^{s_1s_2\cdots s_{a-1} (-s_a) s_{a+1}\cdots s_n} \;
\boldsymbol{\xi}^{s_1}\otimes \boldsymbol{\xi}^{s_2}\otimes \cdots \otimes \boldsymbol{\xi}^{s_{a-1}}\otimes  \boldsymbol{\xi}^{s_{a}} \otimes \boldsymbol{\xi}^{s_{a+1}} \otimes  \cdots \otimes  \boldsymbol{\xi}^{s_n}.
\end{align}
From the first to the second line, we have changed the index $s_a$ to $-s_a$. The final result remains the same since we are summing over all values of $s_a$, which comprise the same list of the values of $-s_a$. 
Analogously, we have
\begin{align}\label{GammaatildePsi}
 \gamma_{\widetilde{a}} \hat{\boldsymbol{\Psi}} & = \sum_{\{s\}}\left(\prod_{b=1}^{a-1}s_b\right)(is_a) \hat{\Psi}^{s_1s_2\cdots s_{a-1} s_a s_{a+1}\cdots s_n} \;
\boldsymbol{\xi}^{s_1}\otimes \boldsymbol{\xi}^{s_2}\otimes \cdots \otimes \boldsymbol{\xi}^{s_{a-1}}\otimes  \boldsymbol{\xi}^{-s_{a}} \otimes \boldsymbol{\xi}^{s_{a+1}} \otimes  \cdots \otimes  \boldsymbol{\xi}^{s_n} \nonumber\\
& =  \sum_{\{s\}}\left(\prod_{b=1}^{a-1}s_b\right) (-is_a)\hat{\Psi}^{s_1s_2\cdots s_{a-1} (-s_a) s_{a+1}\cdots s_n} \;
\boldsymbol{\xi}^{s_1}\otimes \boldsymbol{\xi}^{s_2}\otimes \cdots \otimes \boldsymbol{\xi}^{s_{a-1}}\otimes  \boldsymbol{\xi}^{s_{a}} \otimes \boldsymbol{\xi}^{s_{a+1}} \otimes  \cdots \otimes  \boldsymbol{\xi}^{s_n},
\end{align}
that is, the action of the Dirac matrices on a spinor field results also in a spinor field.
The covariant derivative of a spinor field $\hat{\boldsymbol{\Psi}}$ is defined in terms of the spin connection as
\begin{equation}\label{covD}
\hat{\nabla}_{\alpha}\hat{\boldsymbol{\Psi}}\,=\,\hat{\partial}_{\alpha}\hat{\boldsymbol{\Psi}}
\,-\,\frac{1}{4}\,\hat{\omega}_{\alpha}^{\phantom{\alpha}\beta\varepsilon}\,\gamma_{\beta}\,\gamma_{\varepsilon}\,\hat{\boldsymbol{\Psi}} \,,
\end{equation}
with $\hat{\partial}_{\alpha}$ denoting the partial derivative along the vector field $\hat{\boldsymbol{X}\,}_{\alpha}$.

We are interested in a spinor field $\hat{\boldsymbol{\Psi}}$ such that it belongs to the kernel of the
modified Dirac operator $D^{\hat{\mathcal{F}}}$, that is

\begin{equation}\label{Modified_Dirac_equation}
D^{\hat{\mathcal{F}}}\,\hat{\boldsymbol{\Psi}} = 0\,,
\end{equation}
where $D^{\hat{\mathcal{F}}}$ is the operator
\begin{equation}\label{DOWF}
D^{\hat{\mathcal{F}}} = \hat{D} - \hat{\mathcal{F}} \quad \text{with} \quad \hat{D} = \gamma^{\alpha}\hat{\nabla}_\alpha .
\end{equation}
$\hat{D}$ is the standard Dirac operator and $\hat{\mathcal{F}}$ is an arbitrary inhomogeneous form, namely
\begin{equation}
\hat{\mathcal{F}} = \sum_{p} \hat{\mathcal{F}}_{p} \quad \text{with}\quad \hat{\mathcal{F}}_{p} = \frac{1}{p!}\, \hat{F}_{\alpha_1 \alpha_2 \ldots \alpha_p} \gamma^{\alpha_1} \gamma^{\alpha_2}\ldots  \gamma^{\alpha_p},
\end{equation}
where $\hat{F}_{\alpha_1 \alpha_2 \ldots \alpha_p} = \hat{F}_{[\alpha_1 \alpha_2 \ldots \alpha_p]}$ are components of a $p$-form. This includes the standard massive Dirac equation, the Dirac equation minimally coupled to an electromagnetic field, the Dirac equation in the presence of torsion, as well as more general cases.

We aim to separate the equation \eqref{Modified_Dirac_equation} in its two-dimensional blocks. In order to accomplish this goal, we shall build a suitable ansatz for the spinor components, which agrees with the background's symmetries. Let us assume that the components of the spinor field \eqref{SpinBasis2} take the separable form
\begin{equation}\label{SpinorSeparable}
 \hat{\Psi}^{s_1s_2\cdots s_a\cdots s_n} = \hat{\Psi}_1^{s_1}(x^1,y^1)\,\hat{\Psi}_2^{s_2}(x^2,y^2)\,\cdots\,\hat{\Psi}_a^{s_a}(x^a,y^a)\,\cdots\, \hat{\Psi}_n^{s_n}(x^n,y^n) \,.
\end{equation}
With this form, we can investigate if the modified Dirac equation \eqref{Modified_Dirac_equation} for the general spinor field $\hat{\boldsymbol{\Psi}}$ on a $2n$-dimensional product manifold whose line element is $ds^{2}$ can be reduced to a modified Dirac equation for a spinor field
\begin{equation}
\hat{\boldsymbol{\Psi}}_a = \sum_{s_a}\hat{\Psi}_a^{s_a}(x^a, y^a)\,\boldsymbol{\xi}^{s_a},
\end{equation}
defined on each two-dimensional submanifold whose line element is $ds_a^{2}$.
This allows us to construct spinor fields on $d\hat{s}^{2}$ as a tensor product of spinor fields defined on two-dimensional spaces $d\hat{s}_a^{2}$. Indeed, assuming \eqref{SpinorSeparable}, the spinor field $\hat{\boldsymbol{\Psi}}$ can be written as the following direct product of two-component spinors
\begin{align}
\hat{\boldsymbol{\Psi}}  = \hat{\boldsymbol{\Psi}}_1 \otimes \cdots \otimes\hat{\boldsymbol{\Psi}}_{a-1} \otimes \hat{\boldsymbol{\Psi}}_a \otimes \hat{\boldsymbol{\Psi}}_{a+1} \otimes \cdots \otimes \hat{\boldsymbol{\Psi}}_n .
\end{align}

In order to solve the modified Dirac equation, we need to separate the degrees of freedom of the field, which can be quite challenging in general. Therefore, for the sake of simplicity, let us first perform the separation of the Dirac equation in the case where the fluxes consist of a $ p $ -form field where $p \leq 3 $. Then we will generalize it to an arbitrary $p$-form flux field.

\section{Fluxes of rank $p \leq 3$}\label{Separability_3D}

For the sake of simplicity, let us start with the Dirac equation coupled to a general inhomogeneous flux field up to 3-form given by
\begin{equation}\label{3FF}
\hat{\mathcal{F}} = \hat{F} + \hat{F}_\alpha\gamma^{\alpha} + \frac{1}{2!}\,\hat{F}_{\alpha_1 \alpha_2}\gamma^{\alpha_1}\gamma^{\alpha_2} + \frac{1}{3!}\,\hat{F}_{\alpha_1 \alpha_2 \alpha_3}\gamma^{\alpha_1}\gamma^{\alpha_2}\gamma^{\alpha_3},
\end{equation}
where the term in the sum denoted by $\hat{F}$ in the above equation stands for the $0$-form. In the presence of the above flux, we need to separate the general equation
\begin{equation}\label{MDEW3FF}
\left(\hat{D} - \hat{F}_{\alpha_1} \gamma^{\alpha_1} - \frac{1}{2!}\,\hat{F}_{\alpha_1 \alpha_2}\gamma^{\alpha_1}\gamma^{\alpha_2}- \frac{1}{3!}\,\hat{F}_{\alpha_1\alpha_2 \alpha_3}\gamma^{\alpha_1}\gamma^{\alpha_2}\gamma^{\alpha_3}   \right)   \hat{\boldsymbol{\Psi}} = \hat{F} \,\hat{\boldsymbol{\Psi}} 
\end{equation}
in its two-dimensional blocks. In general, the presence of fluxes spoils the separability of the equation since the flux term act on the spinor fields $\hat{\boldsymbol{\Psi}}$ as a product of different Dirac matrices, each of which acts on $\hat{\boldsymbol{\Psi}}$ according to equations \eqref{GammaaPsi} and \eqref{GammaatildePsi}. For instance, the term $\hat{F}_{abc}\gamma^{a}\gamma^{b}\gamma^{c}$ changes the sign of different indices in the spinor components.  
On the other hand, the covariant derivative \eqref{covD} itself has in its definition a product of Dirac matrices. The question is why then, it does not spoil separability. The reason is that, by the space geometry, the Dirac matrices in the covariant derivative act on the spinor field in a very particular way. Indeed, in the orthonormal frame of vectors \eqref{IPVF}, the only nonvanishing components of the spin connection are $\hat{\omega}_{a\widetilde{a}a}$ e $\hat{\omega}_{\,\widetilde{a}\,\widetilde{a}\,a}$, so that
\begin{align}\label{covDExpandido}
\left\{\begin{matrix}
\hat{\nabla}_{a}\hat{\boldsymbol{\Psi}}&=\,\hat{\partial}_{a}\hat{\boldsymbol{\Psi}}
\,-\,\frac{1}{2}\,\hat{\omega}_{a\widetilde{a}a}\,\gamma_{\,\widetilde{a}}\gamma_{a}\,\hat{\boldsymbol{\Psi}} ,\\ 
\\
\hat{\nabla}_{\widetilde{a}}\hat{\boldsymbol{\Psi}} &=\,\hat{\partial}_{\,\widetilde{a}}\hat{\boldsymbol{\Psi}}
\,-\,\frac{1}{2}\,\hat{\omega}_{\,\widetilde{a}\,\widetilde{a}\,a}\,\gamma_{\,\widetilde{a}}\gamma_{a}\,\hat{\boldsymbol{\Psi}},
\end{matrix}\right.
\end{align}
where by $\hat{\partial}_{a}$ and $\hat{\partial}_{\,\widetilde{a}}$ we mean the derivatives along the
vector fields $\hat{\boldsymbol{X}\,}_{a}$ and $\hat{\boldsymbol{X}\,}_{\widetilde{a}}$. Notice that $\gamma_{\widetilde{a}}$ change the index $s_a$ to $-s_a$, while $\gamma_{\widetilde{a}}$ retrieves it. Thus, in the standard Dirac operator represented by 
\begin{eqnarray}
\hat{D}\,=\,\gamma^{a}\hat{\nabla}_{a}\,+\,\gamma^{\widetilde{a}}\hat{\nabla}_{\widetilde{a}},
\end{eqnarray} 
the only changes in spinor indices that are actually relevant for the separability are those coming from the action of the Dirac matrices $\gamma_{a}$ and $\gamma_{\,\widetilde{a}}$ on the same spinor index $s_a$.

Although the separability of the Dirac equation presented problems in backgrounds with fluxes, we can circumvent them by imposing that the only potentially non-vanishing 2- and 3-form flux terms appear with the following components
\begin{equation}
\hat{F}_{\widetilde{a}a}, \quad \hat{F}_{a\widetilde{b}b}, \quad \text{and} \quad  \hat{F}_{\,\widetilde{a}\,\widetilde{b}\,b},
\end{equation}
all involving the following action on the spinor field
 \begin{align}\label{GammaaGammaatildePsi}
 \gamma_{a}\gamma_{\widetilde{a}}\hat{\boldsymbol{\Psi}}=- \gamma_{\widetilde{a}}\gamma_{a}\hat{\boldsymbol{\Psi}} & = i\sum_{\{s\}}\,s_a\,\hat{\Psi}^{s_1s_2\cdots s_{a-1} s_a s_{a+1}\cdots s_n} \;
\boldsymbol{\xi}^{s_1}\otimes \boldsymbol{\xi}^{s_2}\otimes \cdots \otimes \boldsymbol{\xi}^{s_{a-1}}\otimes  \boldsymbol{\xi}^{s_{a}} \otimes \boldsymbol{\xi}^{s_{a+1}} \otimes  \cdots \otimes  \boldsymbol{\xi}^{s_n}.
\end{align} 
The above imposition can be interpreted as the natural choice for the components of the flux term in order for the Dirac equation in the presence of flux can be satisfactorily separated in the class of spaces considered in this paper.
In fact, the non-vanishing components of the spin connection themselves have a very similar form. 

Now, the Dirac equation takes the following form
\begin{align}\label{Principap=3}
\left[\hat{D} - \left(\hat{F}_{a}\,	\gamma^{a} + \,\hat{F}_{a\widetilde{b}b}\,\gamma^{a}\gamma^{\widetilde{b}}\gamma^{b} \right) - \left( \hat{F}_{\widetilde{a}}\,\gamma^{\widetilde{a}} + \,\hat{F}_{\,\widetilde{a}\,\widetilde{b}\,b}\,\gamma^{\widetilde{a}}\gamma^{\widetilde{b}} \gamma^{b}\right)  \right] \hat{\boldsymbol{\Psi}} =\,\left(\hat{F} + \hat{F}_{\,\widetilde{a}a}\,\gamma^{\widetilde{a}}\gamma^{a}\right)\,\hat{\boldsymbol{\Psi}} ,
\end{align}
With the spinor decomposition in Eq. \eqref{SpinorSeparable}, which is crucial for the integrability of the modified Dirac equation,
and Eqs. \eqref{GammaaPsi} and \eqref{GammaatildePsi} along with the identity \eqref{GammaaGammaatildePsi}, one can show that
\begin{align}
&\sum_{\{s\}} \sum_{a=1}^{n}  \left(\prod_{c=1}^{a-1}s_c\right)\hat{\Psi}_1^{s_1}\,\hat{\Psi}_2^{s_2}\,\cdots\,\hat{\Psi}_{a-1}^{s_{a-1}} \hat{\Psi}_{a+1}^{s_{a+1}}\cdots \hat{\Psi}_n^{s_n} \left[\left(\hat{\partial}_{a} + \frac{1}{2}\,\hat{\omega}_{\,\widetilde{a}a\widetilde{a}} - \hat{F}_{a} + \sum_{\substack{b=1 \\ b\neq a}}^{n}is_b\,\hat{F}_{a\widetilde{b}b}\right) \right. \nonumber\\
&\left. -is_a\left(\hat{\partial}_{\,\widetilde{a}} + \frac{1}{2}\,\hat{\omega}_{a\widetilde{a}a} - \hat{F}_{\,\widetilde{a}} + \sum_{\substack{b=1 \\ b\neq a}}^{n} is_b\,\hat{F}_{\,\widetilde{a}\,\widetilde{b}\,b} \right)  \right] \hat{\Psi}_a^{(-s_a)}\,\boldsymbol{\xi}^{s_1}\otimes \boldsymbol{\xi}^{s_2}\otimes \cdots \otimes \,\boldsymbol{\xi}^{s_{a}}  \otimes  \cdots \otimes \,\boldsymbol{\xi}^{s_{b}} \otimes  \cdots \otimes \boldsymbol{\xi}^{s_n} \nonumber\\
&= \sum_{\{s\}} \left(\hat{F} -\sum_{a=1}^{n} i s_a\,\hat{F}_{\,\widetilde{a}a}\right)\,\hat{\Psi}_1^{s_1}\,\hat{\Psi}_2^{s_2}\,\cdots\,\hat{\Psi}_a^{s_a}\cdots \hat{\Psi}_n^{s_n}\,\boldsymbol{\xi}^{s_1}\otimes \boldsymbol{\xi}^{s_2}\otimes \cdots \otimes \,\boldsymbol{\xi}^{s_{a}}  \otimes  \cdots \otimes\boldsymbol{\xi}^{s_n}  .
\end{align}
The reason for imposing the restrictions in the flux components of the modified Dirac equation is to find out how we can achieve the separability of variables in the spinor field equation.  
In the previous work \cite{Joas2017}, we achieved the separability of the Dirac equation coupled to a gauge field, represented by $\hat{F}_{a}$ and $\hat{F}_{\,\widetilde{a}}$ in our notation, in the class of spaces considered here. For this particular case, in order for the equation to be separable in blocks depending only on the coordinates $\{x^1,\,y^1\}$,  $\{x^2,\,y^2\}$ and so on, the functions $\hat{F}_a$ and $\hat{F}_{\,\widetilde{a}}$ must depend only on the two coordinates $\{x^a,\,y^a\}$  for each choice of $a$
\begin{equation}
\left\{\begin{matrix}
\hat{F}_a = \hat{F}_a(x^a,y^a),\\
\hat{F}_{\,\widetilde{a}} = \hat{F}_{\,\widetilde{a}}(x^a,y^a).	
\end{matrix}\right.
\end{equation}
Beside that, the $0$-form $\hat{F}$ must be a sum over $a$ of functions depending on these pairs of
coordinates, $\{x^a,\,y^a\}$,  for each choice of $a$
\begin{equation}\label{FunctionF}
\hat{F} = \sum_{a=1}^{n} \hat{m}_a \quad \text{with} \quad \hat{m}_a = \hat{m}_a(x^a, y^a) .
\end{equation}
The novelty here are the terms $\hat{F}_{\,\widetilde{a}a}, \hat{F}_{\,a\,\widetilde{b}\,b}$ and $\hat{F}_{\,\widetilde{a}\,\widetilde{b}\,b}$ for 2- and 3-form fluxes. Following the same procedure, these fields should have the following functional dependence
\begin{equation}\label{Am}
\left\{\begin{matrix}
\hat{F}_{\,\widetilde{a}a} = \hat{F}_{\,\widetilde{a}a}(x^a,y^a),\\
\hat{F}_{a\widetilde{b}b} = \hat{F}_{a\widetilde{b}b}(x^a,y^a), \\
\hat{F}_{\,\widetilde{a}\,\widetilde{b}\,b} = \hat{F}_{\,\widetilde{a}\,\widetilde{b}\,b}(x^a,y^a).
\end{matrix}\right.
\end{equation}

These assumptions lead us to the following equation
\begin{equation}\label{Dif.EQ2}
 \sum_{a=1}^{n} \, \left\{\left(\prod_{c=1}^{a-1}s_c\right)\,  \frac{1}{\hat{\Psi}_{a}^{s_{a}}} \,
 \left[\slashed{D}_a^{s_a} +\sum_{\substack{b=1 \\ b\neq a}}^{n} is_b \left(\hat{F}_{a\widetilde{b}b}
 - i s_a \,\hat{F}_{\,\widetilde{a}\,\widetilde{b}\,b} \right)\right]\hat{\Psi}_{a}^{(-s_{a})} - \left(\hat{m}_a - is_a\,\hat{F}_{\,\widetilde{a}a}\right)\right\} = 0 \,,
\end{equation}
where the operator $\slashed{D}_a^{s_a} $ used above is defined as
\begin{equation}\label{Dslash}
 \slashed{D}_a^{s_a}  = \left( \hat{\partial}_a + \frac{1}{2}\,\hat{\omega}_{\,\widetilde{a}a\widetilde{a}} - \hat{F}_{a} \right) - is_a \left( \hat{\partial}_{\,\widetilde{a}} + \frac{1}{2}\,\hat{\omega}_{a\widetilde{a}a} - \hat{F}_{\,\widetilde{a}} \right).
\end{equation}
Note that each term in the sum over $a$ in Eq. \eqref{Dif.EQ2} depends only on the two coordinates $\{x^a, y^a\}$. For the sum of these terms to be zero, each of them must be a constant denoted by $\eta_a^{\{s\}}$, here called a separation constant, with the sum of the constants being null. Then, we find the following set of coupled first-order differential equations for $\hat{\Psi}_a^{s_a}$ and $\hat{\Psi}_a^{(-s_a)}$ as follows
\begin{equation}\label{Dif.EQ3}
  \left(\prod_{c=1}^{a-1}s_c\right) \,
 \left[\slashed{D}_a^{s_a} + \sum_{\substack{b=1 \\ b\neq a}}^{n}\frac{is_b}{3}\left(\hat{F}_{a\widetilde{b}b}
 - i s_a \,\hat{F}_{\,\widetilde{a}\,\widetilde{b}\,b} \right)\right] \hat{\Psi}_{a}^{(-s_{a})} = \left(\eta_a^{\{s\}}+ \hat{m}_a - is_a\,\hat{F}_{\,\widetilde{a}a}\right) \hat{\Psi}_{a}^{s_{a}} \,,
\end{equation}
with the separation constants satisfying the constraint
\begin{equation}\label{Constraint1}
\sum_{a=1}^{n} \eta_a^{\{s\}} = 0.
\end{equation}
In the next section we will generalize this procedure to the most general case, the Dirac equation coupled with an arbitrary flux in the same class of spacetimes.

\section{Arbitrary fluxes}\label{Separability_GeneralCase}

In this section, we shall perform the separation of the Dirac equation when the flux has an arbitrary inhomogeneous form, a combination of zero to p forms. In order to perform this, following the idea presented in the previous section, we can start by separating the even and odd parts of $\hat{\mathcal{F}}$, namely
\begin{equation}
\hat{\mathcal{F}} = \hat{F} + \sum_{m=1}^{n}\frac{1}{(2m)!}\, \hat{F}_{\alpha_1 \alpha_2 \ldots \alpha_{2m}} \gamma^{\alpha_1} \gamma^{\alpha_2}\ldots  \gamma^{\alpha_{2m}} + \sum_{m=0}^{n-1}\frac{1}{(2m+1)!}\, \hat{F}_{\alpha_1 \alpha_2 \ldots \alpha_{2m+1}} \gamma^{\alpha_1} \gamma^{\alpha_2}\ldots  \gamma^{\alpha_{2m+1}}.
\end{equation}
This decomposition will be of great practical relevance in what follows. Then, the modified Dirac equation is written in the form
\begin{equation}\label{MDEWGFF}
\left(\hat{D} - \sum_{m=1}^{n}\frac{1}{(2m)!}\, \hat{F}_{\alpha_1 \alpha_2 \ldots \alpha_{2m}} \gamma^{\alpha_1} \gamma^{\alpha_2}\ldots  \gamma^{\alpha_{2m}} - \sum_{m=0}^{n-1}\frac{1}{(2m+1)!}\, \hat{F}_{\alpha_1 \alpha_2 \ldots \alpha_{2m+1}} \gamma^{\alpha_1} \gamma^{\alpha_2}\ldots  \gamma^{\alpha_{2m+1}}\right)   \hat{\boldsymbol{\Psi}} = \hat{F}\, \hat{\boldsymbol{\Psi}} ,
\end{equation}
where $\hat{F}$ stands for the $0$-form. For this general case, we shall show that the Dirac equation coupled to arbitrary fluxes admits separation whenever the only potentially non-vanishing flux components appear with the following index structure
\begin{equation}\label{General_Structure}
\hat{F}_{\,\widetilde{b_1}b_1 \widetilde{b_2}b_2\cdots \widetilde{b_m}b_m}, \quad  \hat{F}_{a\,\widetilde{b_1}b_1 \widetilde{b_2}b_2\cdots \widetilde{b_m}b_m}, \quad \text{and} \quad  \hat{F}_{\widetilde{a}\,\widetilde{b_1}b_1 \widetilde{b_2}b_2\cdots \widetilde{b_m}b_m} \quad (m = 0, 1, 2, \cdots, n).
\end{equation}
For these terms, Dirac matrices acti on the spinor field as given below
\begin{align}\label{GGtildePsi}
 \gamma_{\,\widetilde{b_1}}\gamma_{\,b_1}\gamma_{\,\widetilde{b_2}}\gamma_{\,b_2}\cdots \gamma_{\,\widetilde{b_m}}\gamma_{\,b_m}\hat{\boldsymbol{\Psi}} =&\, (-i)^{m}\sum_{\{s\}}\,s_{b_1}s_{b_2}\cdots s_{b_m}\,\hat{\Psi}^{s_1s_2\cdots s_{b-1} s_b s_{b+1}\cdots s_n} \;\nonumber\\
 \times & \,\boldsymbol{\xi}^{s_1}\otimes \boldsymbol{\xi}^{s_2}\otimes \cdots \otimes \boldsymbol{\xi}^{s_{b-1}}\otimes  \boldsymbol{\xi}^{s_{b}} \otimes \boldsymbol{\xi}^{s_{b+1}} \otimes  \cdots \otimes  \boldsymbol{\xi}^{s_n}.
\end{align}

Assuming now the spinor components decomposition defined in Eq. \eqref{SpinorSeparable} along with the identity \eqref{GGtildePsi} in Eq. \eqref{MDEWGFF} leads to
\begin{align}
&\sum_{\{s\}} \sum_{a=1}^{n}  \left(\prod_{c=1}^{a-1}s_c\right)\hat{\Psi}_1^{s_1}\,\hat{\Psi}_2^{s_2}\,\cdots\,\hat{\Psi}_{a-1}^{s_{a-1}} \hat{\Psi}_{a+1}^{s_{a+1}}\cdots \hat{\Psi}_n^{s_n} \times \nonumber\\
\times &\,\left[\slashed{D}_a^{s_a} - \sum_{\substack{\{b\}\\ b\neq a}}\sum_{m=1}^{n} (-i)^{m}\,s_{b_1}s_{b_2}\cdots s_{b_m} \left(\hat{F}_{a\,\widetilde{b_1}b_1 \widetilde{b_2}b_2\cdots \widetilde{b_m}b_m}
 - i s_a \,\hat{F}_{\,\widetilde{a}\,\widetilde{b_1}b_1 \widetilde{b_2}b_2\cdots \widetilde{b_m}b_m}\right)\right] \hat{\Psi}_a^{(-s_a)}\nonumber\\
\times &\,\boldsymbol{\xi}^{s_1}\otimes \boldsymbol{\xi}^{s_2}\otimes \cdots \otimes \,\boldsymbol{\xi}^{s_{a}}  \otimes  \cdots \otimes\boldsymbol{\xi}^{s_n}  \nonumber\\
= & \sum_{\{s\}} \left(\hat{F} + \sum_{a=1}^{n}\sum_{\substack{\{b\}\\ b\neq a}}\sum_{m=1}^{n}(-i)^{m} s_{a}s_{b_1}\cdots s_{b_{m-1}}\,\hat{F}_{\,\widetilde{a}\,a\,\widetilde{b_1}b_1\cdots \widetilde{b_{m-1}}b_{m-1}}\right) \hat{\Psi}_a^{s_a}\nonumber\\
\times &\, \hat{\Psi}_1^{s_1}\,\hat{\Psi}_2^{s_2}\,\cdots\,\hat{\Psi}_{a-1}^{s_{a-1}} \hat{\Psi}_{a+1}^{s_{a+1}}\cdots \hat{\Psi}_n^{s_n} \,\boldsymbol{\xi}^{s_1}\otimes \boldsymbol{\xi}^{s_2}\otimes \cdots \otimes \,\boldsymbol{\xi}^{s_{a}}  \otimes  \cdots \otimes\boldsymbol{\xi}^{s_n}  ,
\end{align}
where the differential operator $\slashed{D}_a^{s_a}$ has the same definition as in Eq. \eqref{Dslash}. Again, for the sake of separability, for the general flux case, one needs to assume that all non-zero flux components should depend only on coordinates $\{x^a,  y^a\}$ as summarized below
\begin{align}\label{Restriction_p_form}
\left\{\begin{matrix}
\hat{F}_{\widetilde{a \, }a\,\widetilde{a_1}a_1\cdots \widetilde{a_{m-1}}a_{m-1}} = \hat{F}_{\widetilde{a \, }a\,\widetilde{a_1}a_1\cdots \widetilde{a_{m-1}}a_{m-1}}(x^a,y^a),\\
 \hat{F}_{a\,\widetilde{a_1}a_1\cdots \widetilde{a_{m}}a_{m}} =  \hat{F}_{a\,\widetilde{a_1}a_1\cdots \widetilde{a_{m}}a_{m}}(x^a,y^a),\\
 \hat{F}_{\widetilde{a}\,\widetilde{a_1}a_1\cdots \widetilde{a_{m}}a_{m}} =  \hat{F}_{\widetilde{a}\,\widetilde{a_1}a_1\cdots \widetilde{a_{m}}a_{m}}(x^a,y^a),\\
\hat{F} = \sum_{a=1}^{n} \hat{m}_a(x^a,y^a).
\end{matrix}\right.
\end{align}
Factorizing the spinor components $\hat{\Psi}_1^{s_1}, \hat{\Psi}_2^{s_2}, \cdots, \hat{\Psi}_2^{s_n}$, we conclude that
\begin{align}\label{Principal}
  \left(\prod_{c=1}^{a-1}s_c\right) \,
 \left[\slashed{D}_a^{s_a} -\sum_{\substack{\{b\}\\ b\neq a}}^{n}\sum_{m=1}^{n} (-i)^{m}\,s_{b_1}s_{b_2}\cdots s_{b_m} \left(\hat{F}_{a\,\widetilde{b_1}b_1\cdots \widetilde{b_{m}}b_m}
 - i s_a \,\hat{F}_{\,\widetilde{a}\,\widetilde{b_1}b_1\cdots \widetilde{b_{m}}b_m}\right)\right] \hat{\Psi}_{a}^{(-s_{a})} \nonumber\\
 = \left(\eta_a^{\{s\}}+ \hat{m}_a + \sum_{\substack{\{b\}\\ b\neq a}}^{n}\sum_{m=1}^{n} (-i)^{m}\,s_a s_{b_1}s_{b_2}\cdots s_{b_{m-1}} \,\hat{F}_{\,\widetilde{a}\,a\,\widetilde{b_1}b_1\cdots \widetilde{b_{m-1}}b_{m-1}}\right) \hat{\Psi}_{a}^{s_{a}} \,,
\end{align}
where the new separation constants $\eta_a^{\{s\}}$ satisfy the same constraint \eqref{Constraint1}.

This result is a generalization of the one obtained in \cite{Joas2017}, in which we present the separability of Dirac's equation minimally coupled to a gauge field in even-dimensional manifolds in the form of the direct product of bidimensional spaces.
With this separable form, we should be able to reduce the Dirac equation to first-order differential equations coupled in pairs, namely, an equation involving the component $\hat{\Psi}_a^{-s_a}$ has $\hat{\Psi}_a^{s_a}$ as the source and vice versa. In this case, we can eliminate $\hat{\Psi}_a^{s_a}$ or $\hat{\Psi}_a^{-s_a}$ to obtain a decoupled second-order differential equation for each component $\hat{\Psi}_a^{s_a}$, which is precisely the condition needed for the separability of the generalized Dirac equation.

In practice, for some of the most known spacetimes, the line element depends on the coordinates $t$ and $r$, for instance, with $t$ being a cyclic coordinate of the metric. In this case, the Dirac field equation \eqref{Principal} can be separated and decoupled into a purely radial equation. In the following section, we present an application.

\section{String-Inspired Models in $D = 10$ Spacetime Dimensions}\label{String-Inspired}

In this section, we shall consider the Dirac equation for a massive charged spin-$1/2$ particle propagating in the background of a complex string-inspired model in $D = 10$ spacetime dimension described in \cite{Giribet2018}. The models include black hole solutions charged under both $1$-form and $3$-form fields whose horizons of the base manifold are products of four two-dimensional unit spheres, $S^{2}$. 

For this model, the black hole solution can be covered by coordinates $\{x^{\mu}\}=\{t,r,z_{j},\bar{z}_{j}\}$, where $j \in \{2,3,4,5\}$, such that the line element is given by
\begin{equation}\label{LineElemI}
ds^2 = g_{\mu\nu}dx^{\mu}dx^{\nu}= -f(r)\,dt^2 +\dfrac{1}{f(r)}\,dr^2 + r^{2}\sum_{j=2}^{5}\dfrac{dz_{j}d\bar{z}_{j} }{(1+\frac{1}{4}\kappa\, z_{j}\bar{z}_{j})^{2}} \,.
\end{equation}
In this line element, $f = f(r)$ is a function of the radial coordinate $r$ determined from the fourth-order polynomial equation
\begin{equation}
\alpha^{3} r^{8}\left(g_{0}f^{4}-g_{1}\kappa f^{3}+g_{2}|\kappa| f^{2}-g_{3}\kappa f+g_{4}|\kappa| \right) - g_{5}r^{14} f = T ,
\end{equation}
 with the matter content given by 
 \begin{equation}
 T = -t_{1}(Q_{1})^{2} -t_{2}(Q_{2})^{2}r^{4}+t_{3}\ell_{P}^{8}M r^{7}+t_{4}Q_{m}r^{8}-t_{5}\kappa r^{14}.
 \end{equation}
Here, the parameters $(M, Q_{1}, Q_{2}, Q_m) $ are related to the mass, two electric charges, and one magnetic charge of the black hole, respectively, and $g_{j}, t_{j}$ are specific numerical coefficients \cite{Giribet2018}. The region covered by the coordinates $\{z_{j},\bar{z}_{j}\}$, which is a maximally symmetric manifold, is fully specified by the curvature parameter $\kappa$. It is the $2$-dimensional sphere ($S^{2}$) for $\kappa =1$, the $2$-dimensional hiperbolic space ($\mathbb{H}^{2}$) for $\kappa = -1$ and the $2$-dimensional Euclidean space ($\mathbb{R}^{2}$) for $\kappa=0$.  The detail of the solutions can be consulted in Ref. \cite{Giribet2018}.

Let us consider the case in which the metric tensor of the base manifold is comprised of a product of four unit 2-spheres ($\kappa =1$), which is locally a static black hole solution in the presence of a $1$- and $3$-form fluxes such that their exterior derivatives are, respectively, given by
\begin{equation}
\boldsymbol{\mathcal{F}}_{2} = \frac{Q_{1}}{r^{8}}\,dt\wedge dr, \quad \boldsymbol{\mathcal{F}}_{4} =  \frac{Q_{2}}{r^{4}}\,dt\wedge dr\wedge\sum_{j=2}^{5}\text{vol}(S_{j}^{2}) + Q_{m}\sum_{j<i}^{5}\text{vol}(S_{j}^{2})\wedge\text{vol}(S_{i}^{2}) ,
\end{equation}
with $\text{vol}(S_{j}^{2})$ being the volume form in the $j$th two-sphere. 
Before writing the background metric tensor in the form presented in the previous sections, let us first make a convenient coordinate transformation. Instead of using the complex projective coordinates $\{z_{j},\bar{z}_{j}\}$, let us parameterize the 2-sphere by two spherical angles $\{\theta_j, \phi_j \}$ that are related to complex coordinate $z_j$ as follows
\begin{equation}
\theta_j(z_j) = 2 \,\text{arctan}\left(\frac{|z_j|}{2}\right) \quad \text{and} \quad \phi_j (z_j) = \text{arg}(z_j).
\end{equation}
This transformation map, for instance, the points $z_j=0$ and $z_j=\infty$ in the projective complex domain to $\theta_j =0$ and $\theta_j =\pi$, respectively, which are the points where the spherical coordinate system breaks down. In terms of these coordinates, the line
element \eqref{LineElemI} is written as 
\begin{equation}\label{LineElemII}
ds^2 =g_{\mu\nu}dx^{\mu}dx^{\nu}= -f(r)\,dt^2 +\dfrac{1}{f(r)}\,dr^2 + r^{2}\sum_{j=2}^{5}\left(d\theta_j^2 + \sin^{2}\theta_j\,d\phi_j^2\right)\,,
\end{equation}
and the configurations of the $1$- and $3$-form for such a solution are given by
\begin{equation}
\boldsymbol{\mathcal{F}}_{1} = \frac{Q_{1}}{7\,r^{7}}\,dt, \quad \boldsymbol{\mathcal{F}}_{3} =  \frac{Q_{2}}{3\,r^{3}}\,dt \wedge\sum_{j=1}^{4}\sin\theta_j \, d\theta_j \wedge d\phi_j + Q_{m}\sum_{j<i}^{4} \cos\theta_j\,d\phi_j\wedge \sin\theta_i \, d\theta_i \wedge d\phi_i .
\end{equation}

It is worth recalling that spacetime is spherically symmetric if an action of $SO(3)$ by isometries whose orbits are spacelike two-dimensional spheres exists. Clearly, this is not the case since the angular part of the line element is the direct product of four two-spheres, and therefore, the background has $SO(3)\times SO(3) \times SO(3) \times SO(3)$ symmetry. For each of the four spheres, there exist three independent Killing vectors that generate rotations, namely
\begin{equation}\label{KilllingV}
  \left\{
     \begin{array}{ll}
        \mathbf{K}_{1,j} = \sin\phi_j\, \partial_{\theta_j} + \cot\theta_j \cos\phi_j\,\partial_{\phi_j}\,,\\
       \mathbf{K}_{2,j} = \cos\phi_j\, \partial_{\theta_j} - \cot\theta_j \sin\phi_j\,\partial_{\phi_j} \,,\\
       \mathbf{K}_{3,j} = \partial_{\phi_j}  \,.
     \end{array}
   \right.
\end{equation}
In addition to these Killing vectors, $\mathbf{K}_t = \partial_t$ also generates an isometry. In particular, the existence of the trivial Killing vector fields $\mathbf{K}_t$ and $\mathbf{K}_{3,j}$ translates into the fact that the coefficients of the metric are independents of the coordinates $t$ and $\phi_j$, allowing us to decompose the dependence of the spinor components in these coordinates in the Fourier basis. This will be very useful in order to simplify the radial and angular parts of the Dirac equation.

In the previous section, we showed that the Dirac equation in the presence of arbitrary fluxes is separable in a class of spaces whose metric is the direct product of bidimensional spaces. Clearly, this is not the case here since there is the multiplicative factor $r^{2}$ in front of the line element angular parts. In order to circumvent this limitation, we can factor out the function $r^{2}$ in the line element \eqref{LineElemI} and define 
\begin{equation}
d\hat{s}^2 = \Omega^2\,ds^2 \quad \text{with} \quad \Omega = r^{-1}\, ,
\end{equation}
which is conformally related to our initial line element. This way the new line element becomes
\begin{equation}\label{Conformal_Space}
d\hat{s}^2 = -\frac{f(r)}{r^{2}}\,dt^2 +\dfrac{1}{r^2 f(r)}\,dr^2 + \sum_{j=2}^{5}\left(d\theta_j^2 + \sin^{2}_j\,d\phi_j^2\right) = \sum_{a=1}^{5}d\hat{s}_{a}^{2}\,,
\end{equation}
where
\begin{equation}
d\hat{s}_1^2 =  -\frac{f(r)}{r^{2}}\,dt^2 +\dfrac{1}{r^2 f(r)}\,dr^2, \quad d\hat{s}_j^2 = d\theta_j^2 + \sin^{2}\theta_j\,d\phi_j^2 .
\end{equation}
This conformally transformed space with line element $d\hat{s}^2 $ is a direct product of bidimensional spaces. Indeed,  $d\hat{s}_1^2$ depends just on the coordinates $t$ and $r$, while $d\hat{s}_j^2$ depends on the angular coordinates $\theta_j$ and $\phi_j$.  For this particular class of spacetimes, we are able to obtain the separability of the modified Dirac equation, Eq. \eqref{Principal}. To properly match this separability, we need to introduce a suitable orthonormal frame. Such a frame for the conformal space \eqref{Conformal_Space} is given by  
\begin{equation}\label{Conformal_Frame_BH}
\hat{\boldsymbol{e}}^1 = \frac{i\,\sqrt{f}}{r}\,dt, \quad \hat{\boldsymbol{e}}^{\tilde{1}} = \frac{1}{r\sqrt{f}}\,dr, \quad \hat{\boldsymbol{e}}^{j} = \sin\theta_j\,d\phi_j \quad \hat{\boldsymbol{e}}^{\tilde{j\,}} = d\theta_j .
\end{equation}
The line element can be recast into a simple form in terms of this frame as follows
\begin{equation}
d\hat{s}^{2} = \sum_{a=1}^{5}\left( \hat{\boldsymbol{e}}^{a}\,\hat{\boldsymbol{e}}^{a} + \hat{\boldsymbol{e}}^{\tilde{a}}\,\hat{\boldsymbol{e}}^{\tilde{a}} \right),
\end{equation}
and the $1$- and $3$-form fluxes can be rewritten as 
\begin{equation}
\boldsymbol{\mathcal{F}}_1 = F_{a}\,\hat{\boldsymbol{e}}^{a}  + F_{\tilde{a}}\,\hat{\boldsymbol{e}}^{\tilde{a}}, \quad  \boldsymbol{\mathcal{F}}_3 = F_{a\tilde{b}b}\,\hat{\boldsymbol{e}}^{a} \wedge \hat{\boldsymbol{e}}^{\tilde{b}} \wedge \hat{\boldsymbol{e}}^{b}   + F_{\tilde{a}\tilde{b}b}\,\hat{\boldsymbol{e}}^{\tilde{a}} \wedge \hat{\boldsymbol{e}}^{\tilde{b}} \wedge \hat{\boldsymbol{e}}^{b} ,
\end{equation} 
with their non-vanishing components given by
\begin{equation}\label{p_formComp}
\left\{\begin{matrix}
F_{1} = -\dfrac{i\,Q_1}{7\,r^6\sqrt{f}},\\ 
F_{1\tilde{j\,}j} = -\dfrac{i\, Q_2}{3\,r^2\sqrt{f}} , \quad F_{j\,\widetilde{i}\,i} = Q_m \, \cot\theta_i \, .
\end{matrix}\right.
\end{equation}
Note, for example, that the flux components do not mix the radial and angular coordinates. This is actually a reflection of the fact that the space whose line element is \eqref{Conformal_Space} is formed from the direct product of bidimensional spaces.

Notice that, we could have considered the Dirac equation directly on $ds^2$, which is not a direct product of bidimensional spaces, and then used invariance properties under conformal transformations to relate the relevant quantities concerning the spacetime with line elements $ds^2$ to the quantities defined on $d\hat{s}^2$, a similar procedure done in our previous work \cite{Joas2017}. In doing so, although the resulting equation has the form of the equation that we have been able to separate in the previous section, the components $F_{j\,\widetilde{i}\,i}$ mix the radial and angular coordinates so that we had to consider a vanishing magnetic field in the background. 

Now, let us consider a spin $1/2$ particle of mass $m$ and electric charge $q$, propagating on the background described
by the metric \eqref{Conformal_Space}. Such a particle is represented by a spinor field $\hat{\boldsymbol{\Psi}}$ obeying the following generalized Dirac equation
\begin{equation}\label{Our_Equation_Geral}
\left[\hat{D} - q \left(i\,\mathcal{F}_{1} + \frac{1}{4}\,\mathcal{F}_{3} \right)\right]   \hat{\boldsymbol{\Psi}} = m \,\hat{\boldsymbol{\Psi}} ,
\end{equation}
where $\mathcal{F}_{1}$ and $\mathcal{F}_{3}$ represent the $1$- and $3$-form background fluxes. The factor of $-1/4$ appearing in the above equation was introduced in order to Schr\"{o}dinger-Lichnerowicz type formula is properly satisfied by the Dirac operator in the presence of fluxes of rank $3$ \cite{Hour2010,Agricola2004}. Using the fact that $F_{\tilde{a}} =0$ and $F_{\tilde{a}\tilde{b}b} = 0$, it takes the following form 
\begin{equation}\label{DEDS2}
\left[\hat{D} - q\left(i\,F_{a}\,\gamma^{a} + \frac{1}{4}\,F_{a\tilde{b}b}\,\gamma^{a}\,\gamma^{\tilde{b}}\,\gamma^{b}  \right) \right] \hat{\boldsymbol{\Psi}} = m \,\hat{\boldsymbol{\Psi}} .
\end{equation}

In order to put the above equation in the form of the modified Dirac equation \eqref{Principap=3}, we should set
\begin{eqnarray}\label{use}
\hat{F}_{a} =  i\,q\, F_{a}, \quad \hat{F}_{\widetilde{a}} = 0, \quad \hat{F}_{a\widetilde{a}} = 0, \quad \hat{F}_{a\widetilde{b}b} = \frac{q}{4}\,F_{a\widetilde{b}b},\quad \hat{F}_{\tilde{a}\widetilde{b}b} = 0\,,
\end{eqnarray}
as well as
\begin{equation}
\hat{m}_{1} = m \quad \text{and} \quad \hat{m}_j = 0 .
\end{equation}
Using these relations, we are left with the following field equation
\begin{equation}
\left(\hat{D} - \hat{F}_{a}\,\gamma^{a} - \hat{F}_{a\tilde{b}b}\,\gamma^{a}\gamma^{\tilde{b}}\gamma^{b} \right) \hat{\boldsymbol{\Psi}} = \hat{m}_1 \hat{\boldsymbol{\Psi}},
\end{equation}
which is exactly in the form of the equation that we have been able to separate in the previous section. 

According to the index conventions adopted in the line element \eqref{Conformal_Space}, for $a=1$, we have $d\hat{s}_1^2$, $F_{1}$, and $F_{1\,\widetilde{j}\,j}$ depending only on the coordinates $t$ and $r$, while for $a=j$ we have $d\hat{s}_j^2$, and $F_{j\,\widetilde{i}\,i}$ depending on the angular coordinates $\theta_j$ and $\phi_j$. 
Thus, it is convenient to adopt the following conventions when choosing coordinates
\begin{eqnarray}
x^{1} = t \quad , \quad y^{1} = r \quad , \quad x^{j} = \phi_{j} \quad , \quad y^{j} = \theta_{j} \,.
\end{eqnarray}
In particular, $F_{1\,\widetilde{j}\,j}$ and $F_{j\,\widetilde{i}\,i}$ depend just on the coordinates related to the first index. This is extremely important to validate the separability of the field equation in the background metric, as discussed in Sec \ref{Separability_3D}, since the constraints \eqref{Am} are naturally satisfied. Thus, it follows from Eqs. \eqref{Dslash} and \eqref{Principal} that the radial spinor component $\hat{\Psi}_{1}^{s_1} = \hat{\Psi}_{1}^{s_1}(t, r)$ obeys the following differential equation 
\begin{align}\label{Equacao_Diferencial_R}
\left[\left(  \hat{\partial}_1 + \frac{1}{2} \,\hat{\omega}_{\,\widetilde{1}1\widetilde{1}} - \hat{F}_{1}\right)  - is_1\left( \hat{\partial}_{\,\widetilde{1}} + \frac{1}{2} \,\hat{\omega}_{1\widetilde{1}1} - \hat{F}_{\,\widetilde{1}}  \right)  +\sum_{j=2}^{5} is_j \left(\hat{F}_{1j\widetilde{j}} - i\,s_1 \hat{F}_{\,\widetilde{1}\,\widetilde{j}\,j}
 \right)\right] \hat{\Psi}_{1}^{(-s_{1})} 
 =\, \,\left(\eta_1^{\{s\}}+ \hat{m}_1 \right) \hat{\Psi}_{1}^{s_{1}}  .
\end{align}
The parameters $\eta_1^{\{s\}}$ appearing in the radial equation are constants that depend on angular equation. Each of the angular spinor components $\hat{\Psi}_{j}^{s_{j}} = \hat{\Psi}_{j}^{s_{j}}(\theta_j, \phi_j)$ with $j\geq 2$ satisfies the following differential equation
\begin{align}\label{Equacao_Diferencial_A}
 \left[\left(\hat{\partial}_j + \frac{1}{2} \,\hat{\omega}_{\,\widetilde{j}j\widetilde{j}} - \hat{F}_{j}\right) - is_j\left(\hat{\partial}_{\,\widetilde{j}} + \frac{1}{2} \,\hat{\omega}_{j\widetilde{j}j} - \hat{F}_{\,\widetilde{j}}  \right) + \sum_{i\neq j}^{5} is_i \left(\hat{F}_{j\,\widetilde{i}\,i}
 - i s_j \,\hat{F}_{\,\widetilde{j}\,\widetilde{i}\,i} \right)\right] \hat{\Psi}_{j}^{(-s_{j})} 
 =\, \left(\prod_{a=1}^{j-1}s_a\right) \eta_j^{\{s\}} \,\hat{\Psi}_{j}^{s_{j}} ,
\end{align}

Concerning the frame of $1$-forms $\{\hat{\boldsymbol{e}}^{a}, \hat{\boldsymbol{e}}^{\widetilde{a}}\}$ considered in Eq. \eqref{Conformal_Frame_BH}, the only non-vanishing components of the spin connection and flux field, according to the Eqs. \eqref{p_formComp} and \eqref{use}, are
\begin{equation}\label{Nonnull_Comp}
\left\{\begin{matrix}
\hat{\omega}_{1\widetilde{1}1} = -\,\hat{\omega}_{11\widetilde{1}} = \dfrac{r f' - 2 f}{2\sqrt{f}}, \quad \hat{F}_{1} = -\dfrac{iq\,Q_1}{7\,r^6\sqrt{f}}, \quad \hat{F}_{1\widetilde{j}j} = -\dfrac{i\,q\,Q_2}{12\,r^2\sqrt{f}},\\ 
\hat{\omega}_{j\widetilde{j}j} =  -\, \hat{\omega}_{jj\widetilde{j}} = \cot\theta_j, \quad \hat{F}_{j\,\widetilde{i}\,i} = \dfrac{q\,Q_m}{4}\,\cot\theta_j \quad\text{for}\quad i \neq j\geq 2 ,
\end{matrix}\right. 
\end{equation}
where $f'$ stands for the derivative of $f$ with respect to its variable $r$. Then, by inserting these expressions into Eqs. \eqref{Equacao_Diferencial_R} and \eqref{Equacao_Diferencial_A} we are left with the following differential equations
\begin{align}\label{Equacao_Diferencial_Radial}
&\left[ is_1\left(  \frac{r}{i\sqrt{f}}\, \partial_t + \frac{iq\, Q_1}{7\,r^6 \sqrt{f}} - \frac{iq\,Q_2}{12\,r^2 \sqrt{f}}\sum_{j=2}^{5}s_j\right) + \left(r\sqrt{f}\,\partial_r + \frac{r\,f' - 2f\,}{4\sqrt{f}}\right)  \right] \hat{\Psi}_{1}^{(-s_{1})} 
 =\, \,\left(\kappa_1^{\{s_1,s_j\}}+ is_1 m \right) \hat{\Psi}_{1}^{s_{1}}  ,
\\\label{Equacao_Diferencial_Angular}
&\left[is_j\left( \frac{1}{\sin{\theta_j}}\,\partial_{\phi_j} + \frac{q\,Q_m}{4} \cot{\theta_j}\sum_{i\neq j}^{5}s_i \right)+ \left(\partial_{\theta_j} + \frac{1}{2}\cot{\theta_j}\right)\right] \hat{\Psi}_{j}^{(-s_{j})} 
 =\,\kappa_j^{\{s_1,s_j\}} \hat{\Psi}_{j}^{s_{j}} \quad \text{if} \,\, j\geq 2 ,
\end{align}
where the parameters $\kappa_a^{\{s_1,s_j\}}$ defined as
\begin{equation}
\kappa_a^{\{s_1,s_j\}} = i\left(\prod_{b=1}^{a}s_b\right) \,\eta_a^{\{s\}} \quad \forall \,\, a ,
\end{equation}
are constants that depend on $\{s\} = \{s_1, s_2, s_3, s_4, s_5\}$ for each $a$. 
One needs to solve the above equations for $\hat{\Psi}_{1}^{s_{1}}(t, r)$ and $\hat{\Psi}_{j}^{s_{j}}(\phi_{j}, \theta_{j})$ in the space comprised by the direct product of bidimensional spaces. In the following subsections we intend to analyze the angular and radial parts of the modified Dirac equation in this model.

\subsection{Angular part}

First, let us work out the angular parts of the spinorial field that obey the differential equation \eqref{Equacao_Diferencial_Angular}. Since the coefficients in the equation for $\hat{\Psi}_{j}^{s_{j}}$ do not depend on the coordinate $\phi_j$, stemming from the fact that $\partial_{\phi_j}$ is a Killing vector field of our metric, we can expand the dependence of $\hat{\Psi}_{j}^{s_{j}}$ on the coordinate $\phi$ in the Fourier basis
\begin{equation}
\hat{\Psi}_{j}^{s_{j}}(\phi_{j}, \theta_{j}) = e^{-i\,\omega_j \phi_j}\,\hat{\psi}_{j}^{s_{j}}(\theta_{j}) .
\end{equation}
For notational simplicity, we are omitting the integral over all values of the Fourier frequencies $\omega_j$, which are half-integers since we work with the spin-$1/2$ field. The justification for this comes from the well-known spinorial field property under rotations. 
When a rotation through $2\pi$ is performed in the spinor space, spin-$1/2$ fields are multiplied by $-1$. The way to achieve this is to let the Fourier frequencies $\omega_j$ assume only half-integer values \cite{Abrikosov2002,Camporesi1995}:
\begin{equation}\label{AngularFrequency}
  \omega_j = \pm\, \frac{1}{2}\,,\,\,   \pm\, \frac{3}{2}\,,\,\, \pm\, \frac{5}{2}\,,\,\, \cdots  \;.
\end{equation}
It follows that the field equation \eqref{Equacao_Diferencial_Angular} gives rise to
\begin{align}\label{Equacao_Diferencial_Angular_Phi}
\left[\dfrac{d}{d\theta_j} + \frac{1}{2}\cot{\theta_j} +s_j\left(\frac{\omega_j}{\sin{\theta_j}} + \frac{i\,q\,Q_m}{4} \cot{\theta_j}\sum_{i\neq j}^{5}s_i \right)\right] \hat{\psi}_{j}^{(-s_{j})} 
 =\, \kappa_j^{\{s_1,s_j\}} \hat{\psi}_{j}^{s_{j}} \quad \text{if} \,\, j\geq 2 .
\end{align}
Note that this first-order differential equation mixes the angular components of the spinorial field, $\hat{\psi}_{j}^{+}$
and $\hat{\psi}_{j}^{-}$. In order to separate these components we need to differentiate this equation once more, which, after some algebra, leads to the equation
\begin{align}\label{SecondOrder-Ang}
 & \frac{1}{\sin\theta_j} \frac{d}{d\theta_\ell}\left[ \sin\theta_j \frac{d}{d\theta_j}  \hat{\psi}_{j}^{(-s_j)} \right] + A(\{s\}, \theta_j)\, \hat{\psi}_{j}^{(-s_j)} = 0 ,
\end{align}	
where $A(\{s\}, \theta_j)$ is a function given by
\begin{align}
A(\{s\}, \theta_j) &=  \frac{\left[4 + q^2Q_m^2 \left(\sum_{i \neq j}^{5}s_i\right)^2 \right] \,\cos^2\theta_j }{16\sin^2\theta_j}-\frac{\left(2 + i\,q\,Q_m s_j\sum_{i \neq j}^{5}s_i\right)  s_j\omega_j \cos\theta_j}{2\sin^2\theta_j} \nonumber\\ &- \frac{2+ i\, q\,Q_m s_j\sum_{i \neq j}^{5}s_i + 4\omega_j^2}{4\,\sin^2\theta_j }  -
     \kappa_j^{\{s_1,s_j\}}\kappa_j^{\{s_1,-s_j\}} .
\end{align}

This equation must be supplemented by the requirement of regularity of the fields $\psi_{j}^{s_{j}}$ at the points $\theta_j = 0$ and  $\theta_j = \pi$, where our coordinate system breaks down. These regularity conditions transform the task of solving the latter equation in a Sturm-Liouville problem so that the possible values assumed by the separation constants $\lambda_{j}$ form a discrete set. 

In general, the parameters $\kappa_j^{\{s_1,s_j\}}$ depend of each $2^5$ values of the colletive index $\{s\} = \{s_1, s_2, s_3, s_4, s_5\}$. However, in the case where the black hole magnetic charge $Q_m$ vanish, they cannot depend on all choices of $\{s\}$. Instead, $\kappa_j^{\{s\}}$ depends just on $s_j$, so that we can set $\kappa_j^{\{s_1,s_j\}} = s_j \lambda_j$, where $\lambda_j$ is a constant that does not depend on $\{s\}$. In this case, setting $Q_m=0$, we are left with the following equation 

\begin{align}\label{Equacao_Angular_Dirac}
\left[\dfrac{d}{d\theta_j} + \frac{1}{2}\cot{\theta_j} + \frac{s_j\,\omega_j}{\sin{\theta_j}}\right] \hat{\psi}_{j}^{(-s_{j})} 
 =\, s_j\,\lambda_j\,\hat{\psi}_{j}^{s_{j}} \quad \text{if} \,\, j\geq 2,
\end{align}
which is the Dirac equation on the two-dimensional unit sphere whose eigenvalues $\lambda_j$ are known
\cite{Abrikosov2002,Camporesi1995}. Indeed, the latter equation admits regular analytical solutions on the sphere only when the eigenvalues $\lambda_j$ are nonzero integers
\begin{equation}
\lambda_j = \pm 1, \pm 2, \pm 3, \ldots .
\end{equation}
In this particular case, the parameters $\kappa_1^{\{s\}}$ appearing in the radial equation depend of these eigenvalues and are determined by the following expression \cite{Joas2017}

\begin{equation}
\kappa_1^{\{s\}} = -\lambda_{2} - s_2 \lambda_{3} - s_2 s_3 \lambda_{4} - s_2 s_3 s_4 \lambda_{5} .
\end{equation}
However, in general, the allowed values of this parameter are not known, since for the case $Q_{m} \neq 0$ the solutions of Eq. \eqref{SecondOrder-Ang} are not known. For solving it, one needs to use some semianalytical or numerical methods. It is also possible to look for approximate solutions for the case $Q_{m} \neq 0$ using perturbation techniques, with $Q_m$ being the perturbation parameter,

\subsection{Radial part}

Now, we shall investigate the pair of radial equations in expression \eqref{Equacao_Diferencial_Radial}.
Since  $\partial_t$ is a Killing vector field of our metric, we can expand the time dependence of $\hat{\Psi}_{1}^{s_1}$ in the Fourier basis
\begin{equation}
\hat{\Psi}_{1}^{s_1}(t, r)\,=\, e^{-i\omega t}\,\hat{\Psi}_{1}^{s_1}(r) \,,
\end{equation}
where we are again omitting the integral over $\omega$, for notational simplicity. Inserting the above expression into Eq. \eqref{Equacao_Diferencial_Radial}, we are left with the following differential equation
\begin{align}\label{Equacao_Diferencial_Radial_Phi}
\left[r\sqrt{f}\,\dfrac{d}{dr} +\frac{r\,f' - 2f\,}{4\sqrt{f}} + \dfrac{is_1}{\sqrt{f}}\left(- \omega r + \frac{iq\, Q_1}{7\,r^6} - \frac{iq\,Q_2}{12\,r^2}\sum_{j=2}^{5}s_j\right)   \right] \hat{\Psi}_{1}^{(-s_{1})} 
 = \,\left(\kappa_1^{\{s\}}+ is_1 m \right) \hat{\Psi}_{1}^{s_{1}} .
\end{align}
The form of the radial equation in \eqref{Equacao_Diferencial_Radial_Phi} suggests  that  instead of using the constants $\{\kappa_1^{\{s\}}, m\}$, we can introduce the parameters $\{\zeta^{\{s\}}, \mu^{\{s\}}\}$ defined by
\begin{equation}
\zeta^{\{s\}} = \text{arctan}\left(\frac{m}{\kappa_1^{\{s\}}}\right)  \quad \text{and} \quad \mu^{\{s\}} = \sqrt{\left(\kappa_1^{\{s\}}\right)^2 + m^2} .
\end{equation}
Inverting these expressions, we can prove that the parameters $\{\kappa_1^{\{s\}}, m\}$ are related to the parameters $\{\zeta, \mu\}$ by the following expression
\begin{equation}
\kappa_1^{\{s\}} = \mu^{\{s\}} \, \cos\zeta^{\{s\}} \quad \text{and} \quad m = \mu^{\{s\}}\,\sin\zeta^{\{s\}} .
\end{equation}
In addition to this change of parameters, if we perform yet the field redefinition
\begin{equation}
\hat{\Psi}_{1}^{s_{1}}(r) = e^{-i s_1 \zeta^{\{s\}}/2} \, \Phi_{1}^{s_{1}}(r) ,
\end{equation}
we see that Eq. \eqref{Equacao_Diferencial_Radial_Phi} reduces to 
\begin{align}\label{Equacao_Radial}
\left[\dfrac{d}{dr} + \frac{r\,f' - 2f\,}{4 r f} -\dfrac{is_1}{f}\left(\omega - \frac{iq\, Q_1}{7\,r^7} + \frac{iq\,Q_2}{12\,r^3}\sum_{j=2}^{5}s_j\right)   \right] \Phi_{1}^{(-s_{1})} 
= \,\dfrac{\mu^{\{s\}}}{r\sqrt{f}} \, \Phi_{1}^{s_{1}} .
\end{align}
In order to accomplish solving the radial equation \eqref{Equacao_Radial}, one first needs to decouple the fields $\Phi^{+}_{1}$ and $\Phi^{-}_{1}$,  reducing it into a second-order ordinary differential equation for field. To simplify the general form of this latter equation, we can introduce the functions
\begin{equation}
\left\{\begin{matrix}
A(s_1,s_j,r)  = \dfrac{r\,f' - 2f\,}{4 r f} -\dfrac{is_1}{f}\left(\omega - \dfrac{iq\, Q_1}{7\,r^7} + \dfrac{iq\,Q_2}{12\,r^3}\sum_{j=2}^{5}s_j\right) ,\\ 
B(s_1,s_j,r) = \dfrac{\mu^{\{s\}}}{r\sqrt{f}} .
\end{matrix}\right.
\end{equation}
The final result is that the fields $\Phi^{s_1}_{1}$ satisfy the following decoupled equation
\begin{equation}\label{Second-order-radial}
\dfrac{d^2\Phi_1^{s_1}}{dr^2} + P(\{s\},r)\,\dfrac{d\Phi_1^{s_1}}{dr} +  Q(\{s\},r)\,\Phi_1^{s_1} = 0 ,
\end{equation}
with the functions $P(\{s\},r)$ and $Q(\{s\},r)$ given by
\begin{align}
\left\{\begin{matrix}
\hspace{-2.1cm}P(\{s\},r) =  A(s_1,s_j,r) + A(-s_1,s_j,r) - \dfrac{1}{A(-s_1,s_j,r)}\dfrac{dA(-s_1,s_j,r)}{dr},\\\\ 
Q(\{s\},r) = \dfrac{dA(-s_1,s_j,r)}{dr} + A(s_1,s_j,r)A(-s_1,s_j,r) - B(s_1,s_j,r)B(-s_1,s_j,r) +\\
\hspace{-5.5cm}- \dfrac{A(-s_1,s_j,r)}{B(-s_1,s_j,r)}\,\dfrac{dB(-s_1,s_j,r)}{dr} ,
\end{matrix}\right.
\end{align}
thus achieving the separability of the modified Dirac equation that we were looking for. Just as for the most relevant backgrounds, it is not possible to obtain an exact analytical solution for the radial equation in the background considered here. The function itself $f(r)$ appearing in the metric does not have a closed form in the radial coordinate $r$. Instead, it is determined from the fourth-order polynomial equation. 
On the other hand, a curious point about this class of black holes is that in the case in which quadratic-order term in curvature and fourth-order term in $4$-form flux are included in the action, it is possible to write a dyonic black hole solution in a closed form, with the function $f(r)$ given by \cite{Giribet2018}    
\begin{equation}
f(r) \simeq \dfrac{\sigma}{7} - \dfrac{\Lambda}{36}r^{2} - \dfrac{9 Q_{m}^{2}}{2r^{6}}  - \dfrac{4\pi\ell_{P}M}{r^{7}} + \dfrac{Q_{2}^{2}}{r^{10}} + \dfrac{Q_{1}^{2}}{56r^{14}} +\ldots \,.
\end{equation}
This provides a tractable model for which the physics involved can be rich. 

With the above expression in hand, one can look for solutions for the radial equation \eqref{Second-order-radial}, which can be challenging to attain analytically. One possible approach can be to infer the asymptotic forms of the solution in some domain of interest, for instance, near the infinity and near the values of $r$ for which the function $f$ vanishes, with or without specific charges. Besides that, one can also look for solutions valid in other domains through approximate or numerical methods. However, all these aspects are out of our main focus, which is to show that the separability of the Dirac equation for a class of product manifolds admitting flux fields is possible. These results may motivate the construction of general methods to attain Dirac's equation separability in higher dimensions. A more compact way of establishing the necessary and sufficient conditions that the Dirac equation admits separation is the existence of certain symmetry operators. So, one question that may be raised is under what circumstances the Dirac equation, in the presence of fluxes investigated in this paper, admits a symmetry operator. In this respect, see \cite{Kubiznak2011}.

\section{Conclusion}\label{Conclusion}

In this paper, we have investigated the separability of a generalized Dirac equation with an arbitrary flux field in a higher dimensional class of spaces obtained from the direct product of bidimensional spaces. We showed that the presence of flux fields spoils separability in general. The main result displayed in Eq. \eqref{Principal} has only been possible to reach for the case where non-vanishing flux components appear in the Dirac equation with particular index structure and functional dependence, Eqs. \eqref{General_Structure} and \eqref{Restriction_p_form}. This was extremely important to validate the separation process in bidimensional blocks discussed in Sec \ref{Separability_GeneralCase}. Part of the success in separating the degrees of freedom of the spinor field in the class of spaces considered here was thanks to a suitable representation for the Dirac matrices, along with convenient indices conventions. The work presented here recovers the results previously obtained \cite{Joas2017} considering the standard minimal coupling with a gauge field. 

With the results obtained in this paper, an interesting application consists of studying the propagation of spin-$1/2$ massive charged particle on the the background of a spacetime conformally related to a complex string-inspired model in $D = 10$ dimensions, which includes couplings with $1$-form and $3$-form fields, Eq. \eqref{Conformal_Space}. We show that the modified Dirac equation in this background can be separated by variables into the purely angular parts and the purely radial parts, Eqs. \eqref{SecondOrder-Ang} and \eqref{Second-order-radial}, transforming them into an eigenvalue problem for the separation constants when regularity conditions are applied. In particular, we have argued that the allowed values of the separation constants have yet to be discovered since the solutions of Eqs. \eqref{SecondOrder-Ang} and \eqref{Second-order-radial} are not known, in general. We know them only when the magnetic charges of the background vanish since the angular parts reduce to an eigenvalue problem for the Dirac operator on the two-sphere, whose solutions are known. An exciting feature of this model is the possibility of writing a dyonic black hole solution in a closed form in the general relativity limit. This provides a tractable model for which the physics involved can be quite rich. In particular, we are interested in the quasinormal modes, which are solutions of the field equation satisfying specific boundary conditions generally imposed on the event horizon and the infinity where the field is propagating outwards (see \cite{Joas2018,Joas2020} for example). We plan to investigate this in a future work.


\section{Acknowledgments}

JV thanks Funda\c{c}\~{a}o de Amparo a Ci\^{e}ncia e Tecnologia do Estado de Pernambuco (FACEPE), for the partial financial support. AM acknowledges financial support from the National Council for Scientific and Technological Development - CNPq, Grant No. 309368/2020-0. AM also thanks financial support from the Brazilian agency CAPES and Universidade Federal de Pernambuco Edital Qualis A.


\begin{thebibliography}{9}

\bibitem{Carter1968} B. Carter, \textit{Global structure of the Kerr family of gravitational fields}, Phys. Rev. \textbf{174}, 1559 (1968)

\bibitem{Hughston1973} L. Hughston and P. Sommers, \textit{Spacetimes with Killing tensors}, Commun. Math.
Phys. \textbf{32}, 147 (1973).

\bibitem{Frolov2008} V. P. Frolov and D. Kubiznak, \textit{Higher-dimensional black holes: hidden symmetries
and separation of variables}, Class. Quant. Grav. \textbf{25}, 154005 (2008).

\bibitem{Carlos2014} C. Batista, \textit{Killing-Yano tensors of order $n-1$}, Class. Quant. Grav. \textbf{31},
165019 (2014).

\bibitem{Cartanbook66} E. Cartan, \textit{The Theory of Spinors}, Dover (1966).

\bibitem{Cartan1913} E. Cartan, \textit{Les groupes projectifs qui ne laissent invariante aucune multiplicit\'{e} plane}, Bull. Soc. Math. France \textbf{41} (1913), 53.

\bibitem{Gibbons1993} G. W. Gibbons, R. H. Rietdijk and J. W. van Holten, \textit{SUSY in the sky}, Nucl. Phys. B \textbf{404 }, 42 (1993).

\bibitem{Cariglia2004} M. Cariglia, \textit{Quantum mechanics of Yano tensors: Dirac equation in curved spacetime}, Class. Quant. Grav. \textbf{21}, 1051 (2004).

\bibitem{Nieuwenhuizen1984} P. V. Nieuwenhuizen and N. P. Warne, Integrability Conditions for Killing Spinors, Commun. Math. Phys. 93 , 277 (1984).

\bibitem{Benn_Book1987}Benn,and R. Tucker, \textit{An Introduction to Spinors and Geometry with Applications in Physics}, Adam Hilger, Bristol (1987).

\bibitem{Joas_Book2019} J. Ven\^{a}ncio, \textit{The Spinorial Formalism}. Lambert Academic Publishing, Germany (2019)

\bibitem{Joas_Formalism} J. Ven\^{a}ncio and C. Batista, \textit{Two-Component spinorial formalism using
quaternions for six-dimensional Sspacetimes}, Adv. Appl. Clifford Algebras \textbf{31}, 71 (2021).

\bibitem{Carlos_Classification} C. Batista and B. C. Cunha, \textit{Spinors and the Weyl tensor classification in six
dimensions}, J. Math. Phys. \textbf{54}, 052502 (2013).


\bibitem{Brill1957} D. R. Brill and J. A. Wheeler, \textit{Interaction of Neutrinos and Gravitational Fields}, Rev. Mod. Phys. \textbf{29}, 465 (1957).

\bibitem{Unruh:1973bda}  W.~Unruh, \textit{Separability of the Neutrino Equations in a Kerr Background}, Phys.\ Rev.\ Lett.\  {\bf 31}, 1265 (1973).

\bibitem{Chandrasekhar1979} S. Chandrasekhar,  \textit{The solution of Dirac's equation in Kerr geometry}, Proc. R. Soc. A \textbf{349}, 571 (1976).

\bibitem{Newman1962} E. T. Newman and R. Penrose, \textit{An Approach to Gravitational Radiation by a Method of Spin Coefficients}, J. Math. Phys. (N.Y.) \textbf{3}, 566 (1962).

\bibitem{Page:1976jj}  D.~N.~Page, \textit{Dirac Equation Around a Charged, Rotating Black Hole},   Phys.\ Rev.\ D {\bf 14}, 1509 (1976).

\bibitem{Walk-Pen1970} M. Walker and R. Penrose, \textit{On quadratic first integrals of the geodesic equations for type \{22\} spacetimes}, Commun. Math. Phys. \textbf{18}, 265 (1970).

\bibitem{Wu2009} S. Q. Wu, \textit{Separability of a modified Dirac equation in a five-dimensional rotating, charged black hole in string theory}, 	Phys. Rev. D \textbf{80}, 044037 (2009).

\bibitem{Semiz1992} I. Semiz, \textit{Dirac equation is separable on the dyon black hole metric}, Phys. Rev. D \textbf{46}, 5414(1992).

\bibitem{Dudley1979} A.  L. Dudley  and  J. D. Finley,  \textit{Covariant perturbed wave equations in  arbitrary type-D 
backgrounds}, J. Math. Phys.  \textbf{20}, 311 (1979).

\bibitem{Joas2017} J. Ven\^ancio and C. Batista, \textit{Separability of the Dirac equation on backgrounds that are the direct product of bidimensional spaces}, Physical Review D \textbf{95}, 084022 (2017).

\bibitem{Joas2018} J. Ven\^{a}ncio and C. Batista, \textit{Quasinormal modes in generalized Nariai spacetimes}, Phys. Rev. D \textbf{97}, 105025 (2018).

\bibitem{Benenti1997} S. Benenti, \textit{Intrinsic characterization of the variable separation in the Hamilton Jacobi equation}, J. Math. Phys. \textbf{38}, 6578 (1997).

\bibitem{Demianski1980} M. Demianski and M. Francaviglia, \textit{Separability structures and KillingYano tensors in vacuum type-D space-times without acceleration}, Int. J. Theor. Phys. \textbf{19}, 675 (1980).

\bibitem{Fatibene2002} L. Fatibene, M. Ferraris, M. Francaviglia and R. G. McLenaghan, \textit{Generalized symmetries in mechanics and field theories}, J. Math. Phys. \textbf{43}, 3147 (2002).

\bibitem{Mohammadi2013} S. S. Gousheh, A. Mohammadi, and L. Shahkarami, \textit{Casimir energy for a coupled fermion-kink system and its stability}, Phys. Rev. D \textbf{87}, 045017 (2013).

\bibitem{Mohammadi2015}A. Mohammadi, E. R. Bezerra de Mello and A. A. Saharian, \textit{Induced fermionic currents in de Sitter spacetime in the presence of a compactified cosmic string}, Class. Quantum Grav. \textbf{32}, 135002 (2015).

\bibitem{Mohammadi2020} E. A. F. Bragan\c{c}a, E. R. Bezerra de Mello, and A. Mohammadi, \textit{Induced fermionic vacuum polarization in a de Sitter spacetime with a compactified cosmic string}, Phys. Rev. D \textbf{101}, 045019 (2020).

\bibitem{Charlton1997} I. M. Benn and P. Charlton, \textit{Dirac symmetry operators from conformal Killing-Yano tensors}, Class. Quant. Grav. \textbf{14}, 1037 (1997).

\bibitem{Carter1979} B. Carter and R.G. McLenaghan, \textit{Generalized total angular momentum operator for the Dirac equation in curved space-time}, Phys. Rev. D \textbf{19}, 1093 (1979).

\bibitem{Cariglia2011} M. Cariglia, P. Krtou$\check{\text{s}}$, D. Kubiz$\check{\text{n}}$\'{a}k, \textit{Dirac equation in Kerr-NUT-(A)dS spacetimes: Intrinsic characterization of separability in all dimensions},  Phys. Rev. D \textbf{84}, 024008 (2011).

\bibitem{Oota2008} T. Oota and Y. Yasui, \textit{Separability of Dirac equation in higher dimensional Kerr-NUT-de Sitter spacetime}, Phys. Lett. B \textbf{659}, 688 (2008).

\bibitem{Freedman2012} D. Z. Freedman and A. van Proeyen, \textit{Supergravity}, Cambridge University Press (2012).

\bibitem{Brown2014B} A. R.Brown, A. Dahlen, A. and A. Masoumi, \textit{Flux compactifications on $(S^{2})N$}. Physical Review D \textbf{90}, 045016 (2014).

\bibitem{Brown2014A} A. R. Brown and A. Dahlen, \textit{Spectrum and stability of compactifications on product manifolds}, Phys. Rev. D \textbf{90}, 044047 (2014).

\bibitem{Martucci2005} L. Martucci, J. Rosseel, D. V. den Bleeken and A. V. Proeyen, \textit{Dirac actions for D-branes on backgrounds with fluxes}, Class. Quant. Grav. \textbf{22}, 2745 (2005).

\bibitem{Tripathy2005} P. K. Tripathy and S. P. Trivedi, \textit{D3 Brane Action and Fermion Zero Modes in Presence of Background Flux}, JHEP \textbf{0506}, 066 (2005).

\bibitem{Grana2002} M. Gra\~{n}a ,\textit{D3-brane action in a supergravity background: the fermionic story}, Phys. Rev. D \textbf{66}, 045014 (2002).

\bibitem{Cariglia2012} M. Cariglia, \textit{Hidden symmetries of Eisenhart-Duval lift metrics and the Dirac equation with flux}, Phys.  Rev. D \textbf{86}, 084050 (2012).

\bibitem{Kubiznak2011} D. Kubiz$\check{\text{n}}$\'{a}k, C. M. Warnick and P. Krtou$\check{\text{s}}$, \textit{Hidden symmetry in the presence of fluxes}, Nucl. Phys. B \textbf{844}, 185 (2011).

\bibitem{Giribet2018}  G. Giribet, M. Lagos, J. Oliva, and A. Vera, \textit{$D = 10$ dyonic black holes in string inspired models}, Phys. Rev. D \textbf{98},  064022 (2018).

\bibitem{Hour2010} T. Houri, D. Kubiz$\check{\text{n}}$\'{a}k, C. M. Warnick, and Y. Yasui, \textit{Symmetries of the Dirac operator with skew-symmetric torsion}, Class. Quant. Grav. \textbf{27}, 185019 (2010).

\bibitem{Agricola2004} I. Agricola and T. Friedrich, \textit{On the holonomy of connections with skew-symmetric torsion}, Math. Ann. \textbf{328},711 (2004).

\bibitem{Abrikosov2002} A. A. Abrikosov, Jr., \textit{Fermion states on the sphere $S^{2}$}, Int. J. Mod. Phys. A \textbf{17}, 885 (2002).

\bibitem{Camporesi1995} R.~Camporesi and A.~Higuchi, \textit{On the Eigen functions of the Dirac operator on spheres and real hyperbolic spaces},  J.\ Geom.\ Phys.\  {\bf 20} 1, (1996).



\bibitem{Joas2020} J. Ven\^ancio and C. Batista, \textit{Spin-2 quasinormal modes in generalized Nariai spacetimes}, Phys. Rev. D \textbf{101}, 084037 (2020).



\end{thebibliography}
\end{document}